\documentclass[preprint,12pt,authoryear]{elsarticle}

\newcommand{\rosat}{{\it ROSAT}}

\newcommand{\swift}{{\it Swift}}

\newcommand{\tess}{\textit{TESS}}
\newcommand{\crts}{\textit{CRTS}}
\newcommand{\wasp}{\textit{SuperWASP}}
\newcommand{\asas}{\textit{ASAS-SN}}
\newcommand{\gaia}{{\it Gaia}}
\newcommand{\asass}{{\it ASAS-3}}

\usepackage{soulutf8} 

\usepackage{amssymb}
\usepackage{orcidlink}
\usepackage{soul}
\usepackage{changes}
\usepackage{hyperref}
\usepackage{footnote}
\usepackage{natbib}

\journal{New Astronomy}

\begin{document}

\begin{frontmatter}



\title{The contact binary system TYC~7275-1968-1 as seen by optical, UV and X-ray observations}

\author[label1,label2]{I. J. Lima} 
\ead{isabellima01@gmail.com}
\affiliation[label1]{organization={CONICET-Universidad de Buenos Aires, Instituto de Astronomía y Física del Espacio (IAFE)},
               addressline={Av. Inte. Güiraldes 2620},
               city={Buenos Aires},
               postcode={C1428ZAA},
               country={Argentina}}
\affiliation[label2]{organization={Universidad Nacional de San Juan, Facultad de Ciencias Exactas, Físicas y Naturales},
             addressline={Av. Ignacio de la Roza 590 (O), Complejo Universitario "Islas Malvinas"},
             city={Rivadavia},
             postcode={J5402DCS},
             state={San Juan},
             country={Argentina}}

\author[label3]{A. C. Mattiuci}
\affiliation[label3]{organization={Instituto Nacional de Pesquisas Espaciais (INPE/MCTI)},  
            addressline={Av. dos Astronautas, 1758}, 
            city={São José dos Campos}, 
            state={São Paulo}, 
            country={Brazil}}

\author[label4]{G. J. M. Luna}
\affiliation[label4]{organization={CONICET-Universidad Nacional de Hurlingham},
            addressline={Av. Gdor. Vergara 2222}, 
            city={Villa Tesei},
            state={Buenos Aires},
            country={Argentina}}

\author[label5]{A. S. Oliveira}
\affiliation[label5]{organization={IP\&D, Universidade do Vale do Paraíba}, 
            addressline={Av. Shishima Hifumi, 2911,
            Urbanova, 12244-000},  
            city={São José dos Campos}, 
            state={São Paulo}, 
            country={Brazil}}

\author[label3]{C. V. Rodrigues}
            
\author[label5]{N. Palivanas}

\author[label2]{N. E. Nuñez}

\begin{abstract}  

We present an analysis of publicly available X-ray and optical observations of \mbox{TYC~7275-1968-1}, a contact binary, red nova progenitor candidate. The long optical time series of \asass, \wasp, \crts, \gaia, \asas, and \tess\ enabled us to improve its orbital period to 0.3828071~$\pm$~0.0000026~d.
We show the presence of an X-ray and UV source associated with \mbox{TYC~7275-1968-1} from Neil Gehrels \swift~Observatory, that was previously assumed to be the counterpart of CD~-36~8436 (V1044~Cen), a symbiotic star located 22~arcsec from the red nova candidate. The X-ray data indicate the presence of a region with a temperature of $kT$~=~0.8$^{+0.9}_{-0.1}$~keV and a luminosity of 1.4$^{+0.1}_{-0.2}$~$\times$~10$^{31}$~erg~s~$^{-1}$ in the range 0.3~--~10~keV. 
The detection of X-rays and modulated UV emission suggests that both components of the binary are chromospherically active.
\end{abstract}




\end{frontmatter}


\section{Introduction}

Red novae and luminous red novae are considered the result of mergers of contact binary systems, which can eject the common envelope 
\citep[e.g.,][]{Pastorello_2019}. The resulting optical transients show luminosities between those of classical novae and supernovae, and they have atypical red colors.
\cite{Kochanek_2014} predicted an occurrence rate for a merger of about $\sim 0.1$ to $\sim 0.03$~yr$^{-1}$ for bright red novae in our Galaxy, such as V838~Mon and V1309~Sco. 
Indeed, nova V838~Mon (M$_{V}$~$<$~7~mag) is proposed to be the result of a stellar merger, but its progenitor still remains unidentified. 
So far, V1309~Sco (M$_{V}$~$<$~10~mag) is the only red nova unambiguously originated from the merging of a low mass-ratio contact binary system that was known before the red nova event \citep{Mason_2010}. 

\cite{Wadhwa_2022b} proposed to identify potential red-nova progenitors based on the temperature of the components, mass ratio ($q$), and amplitude of the light curve determined by some geometrical parameters such as the degree of contact and inclination. The low-mass-contact-binary systems \mbox{(0.6~M$_{\odot}<$~M$_{1}<~$1.4~M$_{\odot}$)} candidates to red-nova progenitors have a narrow period range from 0.27~d to 0.4~d with a peak near 0.35~d (see Figure~3 of \citealt{Wadhwa_2022b}). These systems tend to have a dynamically unstable orbit. However, it is possible that periods above 0.5~d can also present orbital instabilities if the mass ratio is extremely low ($q~<~0.3$, see Figure~63 of \citealt{Kobulnicky_2022b}).

\mbox{TYC~7275-1968-1} (from now on TYC~7275), also catalogued as CRTS J131559.4-370018, was first identified as an eclipsing binary by the Catalina Real-Time Transient Survey (\crts) with an orbital period of 0.382807~d and a mean magnitude of 11.37~mag with a dispersion of few tenths of mag in V band \citep{Drake_2017}. It is a low-mass binary and a candidate to red-nova progenitor with an inclination of 69.8~$\pm$~1.5$^{\circ}$, $q~=~$0.075, and deep contact (up to 95\%), indicating that the temperatures of the components are similar \citep{Wadhwa_2022a}. 
X-ray emission was detected by \rosat~at less than 20 arcsec from TYC~7275, with an exposure time of 351~s and a count rate of 3.18~$\times$~10$^{-2}$~counts~s$^{-1}$ \citep{Voges_2000}. The measured flux was 3.36~$\times$~10$^{-13}$~erg~cm$^{-2}$~s$^{-1}$ and the luminosity was estimated to be 4.27~$\times$~10$^{30}$~ergs~s$^{-1}$ at a quoted distance of 326.7~$\pm$~60~pc \citep{Wadhwa_2022a}. 

In this work we report the analysis of \asass, \wasp, \crts, \gaia, \asas, \tess~and \swift~data of TYC~7275. 
In Section~\ref{sec:optical}, we present the optical data, while Section~\ref{sec:xray} presents the \swift\ data.
Section~\ref{sec:time} and Section~\ref{sec:xrays} show the results of optical time series and X-ray spectral analysis, respectively.  
The discussion and conclusion are summarized in Section~\ref{sec:discussion}.

\section{Observations}

\subsection{Optical data} \label{sec:optical}

\begin{figure*}
\centering
\includegraphics[scale=0.45]{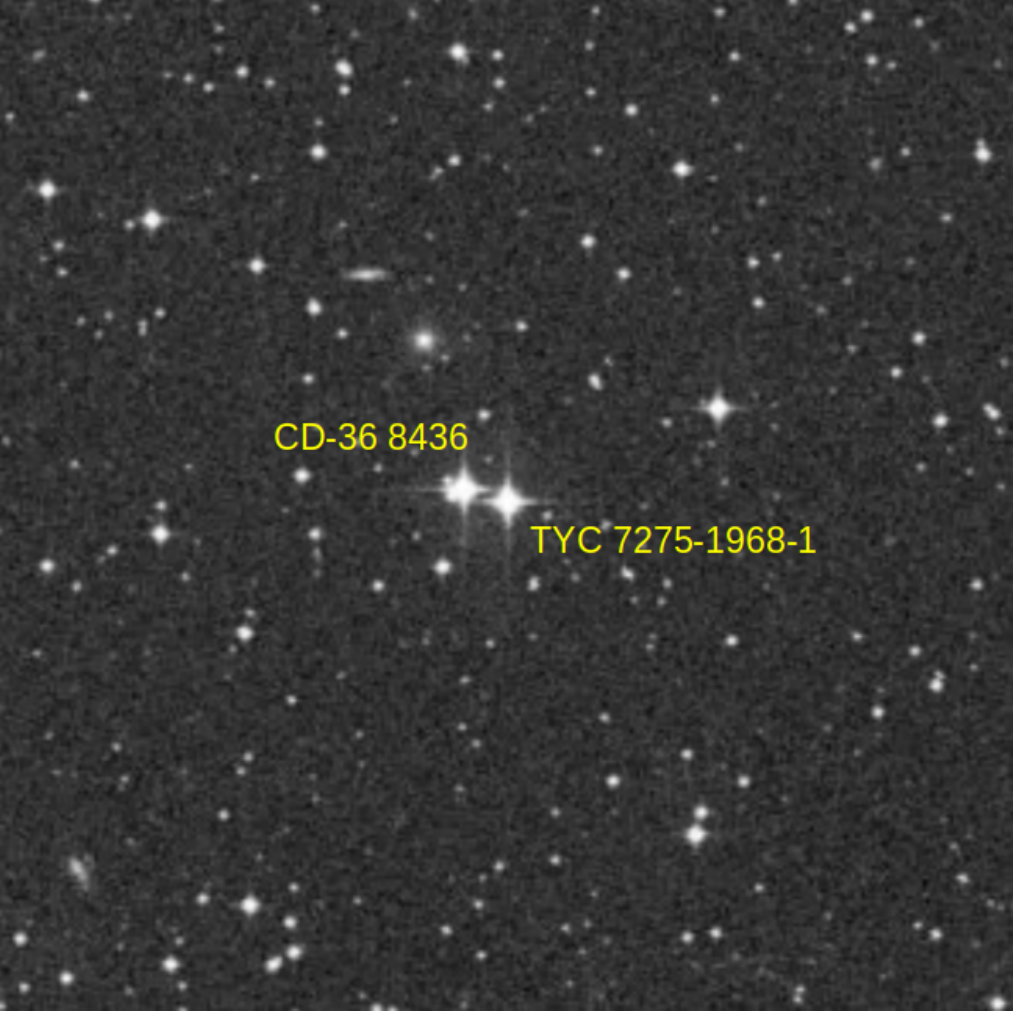}
\caption{Optical image centered on TYC~7275-1968-1 with 8.5~$\times$~8.5 arcmin from DSS \citep{Lasker_1990}. North is up and East is left.}
\label{dss_optical}
\end{figure*}

\begin{figure*}
\begin{center} 
\includegraphics[scale=0.4,angle=0]{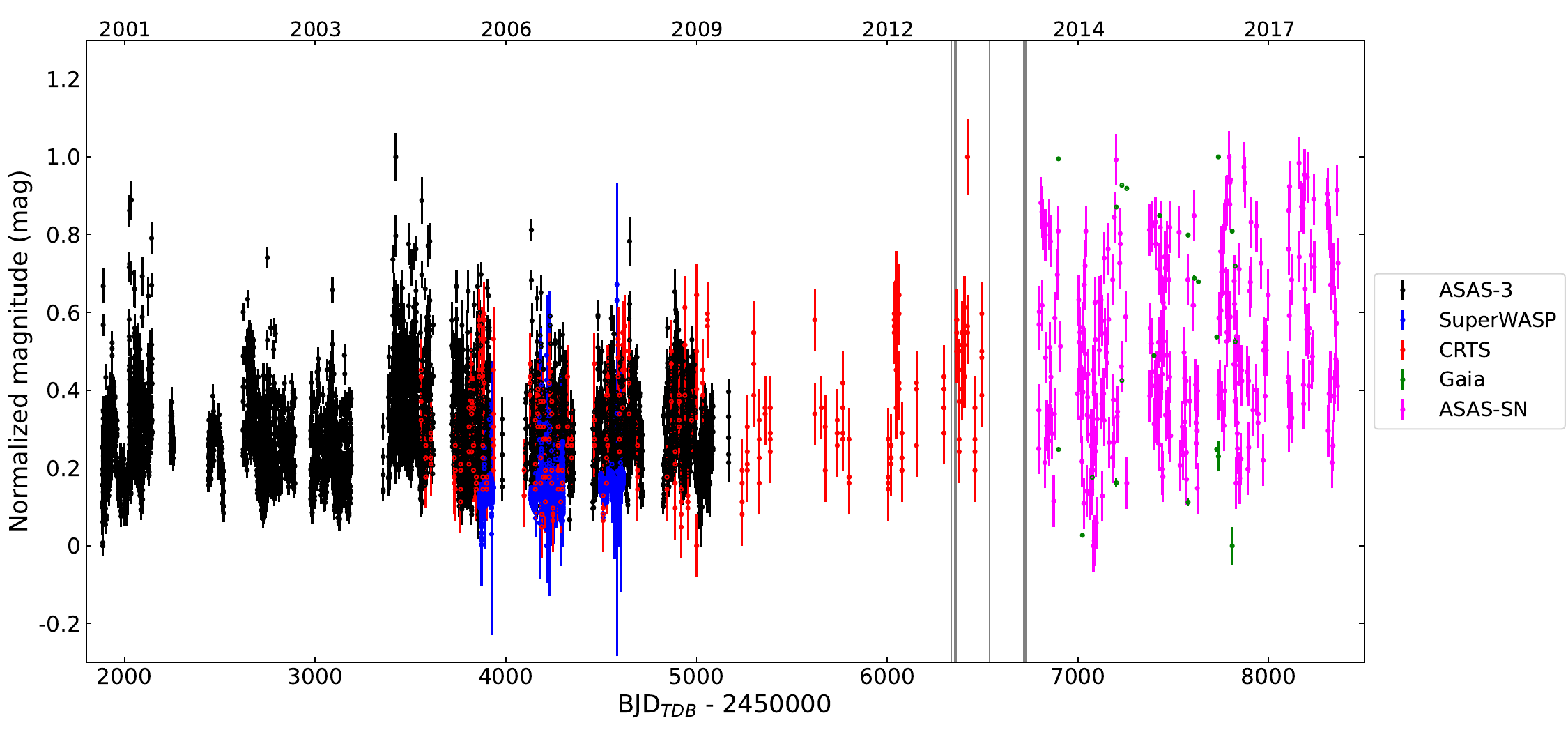}\\
\includegraphics[scale=0.4,angle=0]{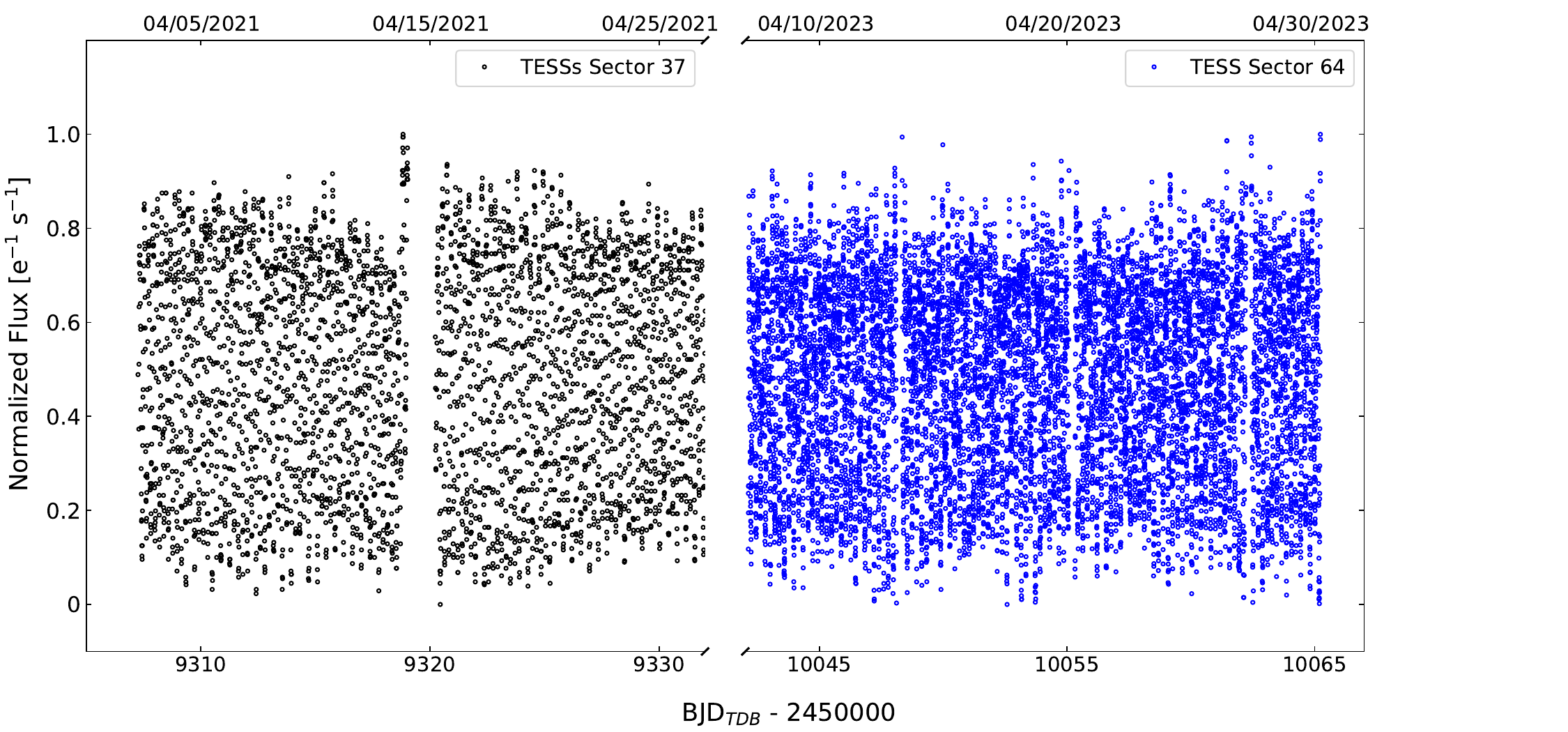}\\ 
\caption{Top panel: Historical light curves of TYC~7275-1968-1 from \asass, \wasp, \crts, \gaia, and \asas. The timings of the \swift\ observations are represented by vertical lines. The bottom panel shows the \tess-PDCSAP flux for Sector 37 and the FFIs extracted flux for Sector 64 (bottom). The gaps in the \tess\ light curves are due to interruption of observations for downloading data to Earth.} 
\label{fig:tess}
\end{center}
\end{figure*}

The Digitized Sky Survey \citep[DSS,][]{Lasker_1990} optical image of the field of TYC~7275 is shown in Figure~\ref{dss_optical}. TYC~7275 and the symbiotic star CD~-36~8436 have similar magnitudes in the V band, around 11~mag. Asides the {\it CRTS} photometric monitoring, TYC~7275 was also observed by the {\it ASAS-3} and {\it ASAS-SN} surveys in V band \citep{pojma_2001, Shappee_2014}, where it is cataloged as ASASSN-V J131559.62-370018.8, and by the {\it SuperWASP} survey in a broad visible band, where it is registered as 1SWASP J131559.55-370017.7, and by {\it Gaia} in the G passband, which covers a wavelength range from the near ultraviolet ($\sim$330~nm) to the near infrared ($\sim$1050~nm), where it is identified as Gaia~DR3~6165817018204434816. The long-term light curves are presented in the upper panel of Figure~\ref{fig:tess}. We converted the time HJD$_{UTC}$ from \asass, \asas, \wasp, and \crts\ surveys and BJD$_{TCB}$ from \gaia\ into BJD$_{TDB}$ using routines in {\it Astropy} and Eastman's time utilities\footnote{See Eastman's website in \url{https://astroutils.astronomy.osu.edu/time/}} \citep{Eastman_2010, astropy_2013}. The time standard given by \tess\ is already BJD$_{TDB}$. We also normalized the magnitude or flux in order to enable a direct comparison. The normalization converted the observed values of each dataset to the interval between 0 and 1.
The average magnitudes and colors are given in Table~\ref{tab:mag_survey} for all available surveys including {\it Tycho-2} and {\it 2MASS}.

\begin{table}
\caption{Photometric measurements and color indices of TYC~7275 from multiple surveys.}  
\smallskip
\centering
\label{tab:mag_survey}
\scalebox{0.6}{
\begin{tabular}{cccccccccc} 
\hline\hline  
\noalign{\smallskip}
Surveys & Band  & Magnitudes & (Bp-Rp)$_{Gaia}$ &(B-V)$_{Tycho-2}$ & (J-H)$_{2MASS}$ & (H-K)$_{2MASS}$ & (J-K)$_{2MASS}$ & Reference \\
& (nm) & & (mag) & (mag) & (mag) & (mag) & (mag) & \\
\noalign{\smallskip}
\hline
\noalign{\smallskip}  
ASAS-3 & 551 & 10.73 & $-$ & $-$ & $-$& $-$& $-$& \cite{pojma_2001}\\
SuperWASP & 400~-~700 & 10.66 & $-$ & $-$ & $-$& $-$& $-$& \cite{Pollacco_2006}\\
CRTS & 540 & 11.37 & $-$ & $-$ & $-$& $-$& $-$& \cite{Drake_2009}\\
Gaia & 330~-~1050 & 11.41 & 0.86 & $-$& $-$& $-$& $-$& \cite{gaia_2020}\\
ASAS-SN & 551 & 11.46 & $-$ & $-$ & $-$& $-$& $-$& \cite{Shappee_2014}\\
Tycho-2 & 425 & 12.18 & $-$ & 0.69 & $-$ & $-$& $-$&\cite{Hog_2000}\\
Tycho-2 & 525 & 11.49 & $-$ & $-$ & $-$& $-$& $-$& \cite{Hog_2000}\\
2MASS &  1250 & 10.28 & $-$ & $-$& 0.29 & $-$ & $-$&\cite{Cutri_2003}\\
2MASS & 1650 & 9.99  & $-$ & $-$& $-$& 0.08 & $-$& \cite{Cutri_2003}\\
2MASS &  2170 &  9.91 & $-$ & $-$& $-$& $-$& 0. 37& \cite{Cutri_2003} \\
\hline \noalign{\smallskip} \end{tabular} 
}
\end{table}

\tess\ observed TYC~7275 in Sector 37 during 25.32~days, between 2021 April 2 and 2021 April 28. Recently, the source has also been observed in Sector 64 during 2023 April 6 to May 3, totalizing 26.29~days of data. Both SAP (Simple Aperture Photometry) and PDCSAP (Pre-search Data Conditioning SAP) light curves from Sector 37 are available from the Quick-Look Pipeline (QLP), with a 10~min cadence. We used the PDCSAP flux for Sector 37, and extracted the Sector 64 light curve from 2~min cadence Full Frame Images (FFIs) using the {\it Lightkurve} Python package \citep{lightkurve}. The \tess\ detectors have a plate scale of 21 arcsec per pixel \citep{Ricker_2015}, so each pixel is large enough to contain multiple stars. All \tess\ light curves (SAP, PDCSAP, and ours) were produced with an aperture mask that blends the fluxes from TYC~7275 (with 11.46 \gaia\ magnitude), from the symbiotic star CD-36~8436 (Gmag~=~9.81) and also from several other fainter sources (with $14~-~20$ Gmag). The normalized \tess\ light curves of TYC~7275 are shown in the lower panel of Figure~\ref{fig:tess}, in which the long-term trend of the fluxes is removed. Our analysis of \tess\ data is based in these light curves.

\begin{figure}
\begin{center}
\includegraphics[scale=0.4,angle=0]{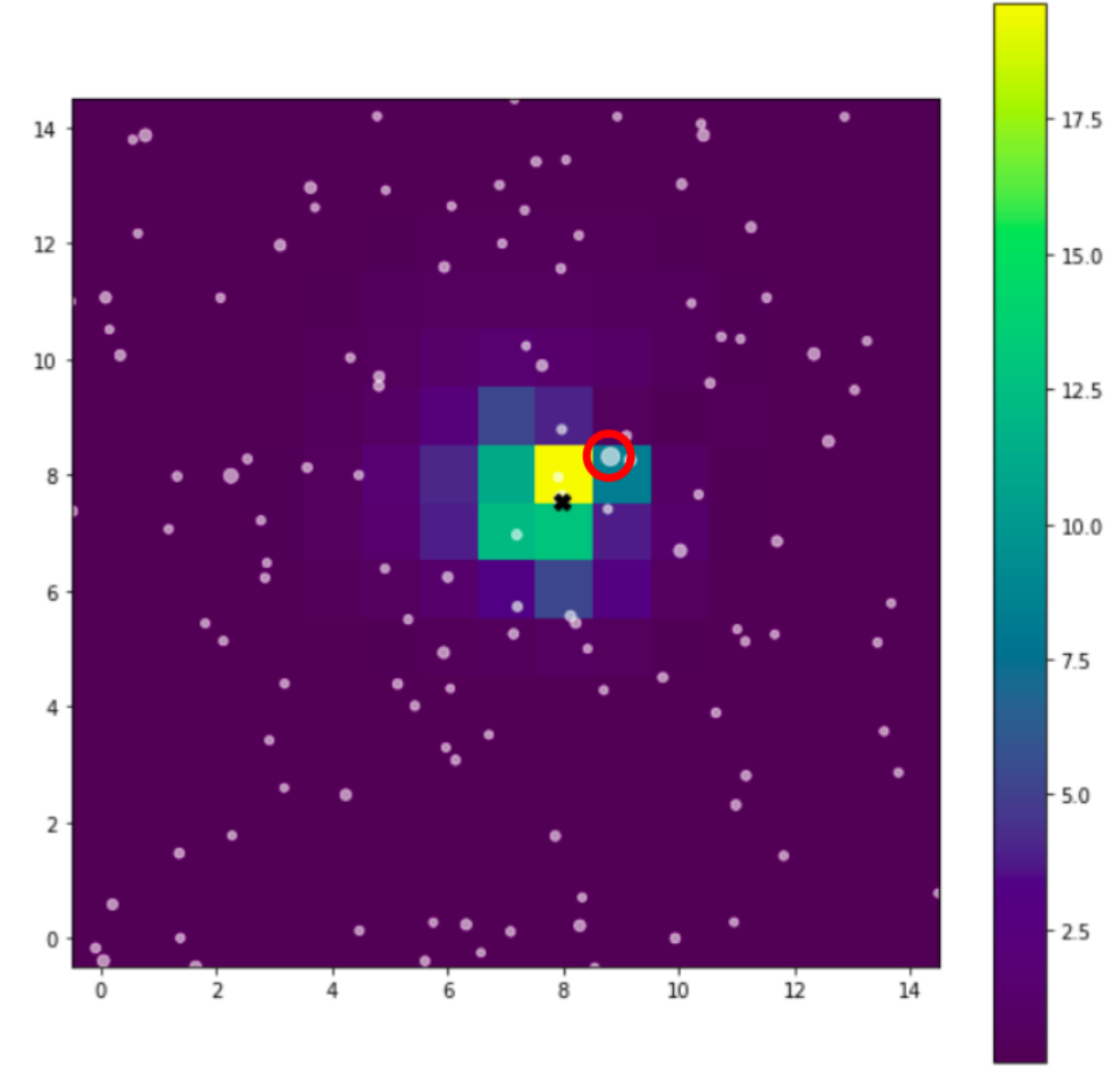}
\caption{\tess\ Target Pixel File of the field of TYC~7275. The location of TYC~7275 is indicated by the black cross. The red ellipse marks the location of the symbiotic star CD-36~8436. \gaia\ sources brighter than 21~mag are represented by the white dots. The location of the variability signal from \texttt{TESS$\_$Localize} coincides with the \gaia\ coordinates of TYC~7275 with 91.8\% confidence. 
The lateral bar shows the flux color scale in $e^{-}/s$. 
}
\label{fig:tess_localize}
\end{center}
\end{figure}

We applied the open-source Python package \texttt{TESS$\_$Localize} \citep{Higgins2022software} to the \tess\ Target Pixel File (TPF) data of TYC~7275. \texttt{TESS$\_$Localize} can locate a variable source in a crowded field to better than one fifth of a pixel, using prior knowledge of the source's periodicity. We used the period of 0.382807~d already known from \cite{Drake_2017}, and also an harmonic of that period (0.191403~d) that is present in the \tess\ data (see Sect.~\ref{sec:time}). The code calculates the relative probability that the source is correlated with the positions of stars from the \gaia\ archive using the pixel response function (PRF\footnote{See \url{https://heasarc.gsfc.nasa.gov/docs/tess/observing-technical.html\#point-spread-function}}). For this set of data, the code confirmed the position of the variable source at the exact \gaia\ coordinates of TYC~7275 (see Figure~\ref{fig:tess_localize}) with a relative likelihood of 91.8\%.

The light curves of TYC~7275 provided by the \crts\ and by \tess\ are shown in Figure~\ref{fig:tess} using daily average magnitudes. We notice that in the \tess\ band, the magnitude increases about 0.05~mag during the 26 days of observation and this behavior also occurs when comparing it with \crts\ data. This long-term variation can not be associated with the orbital motion.
The \tess\ detector passband covers a wide red wavelength range, between 600 and 1000~nm, centered on the Cousins I band. 
While, the \crts\ observations were obtained during 8 years in V band, ranging from 11.2 to 11.6 magnitudes. The ID source in \crts\ is 3037080042793.

\subsection{Neil Gehrels \swift\ data} \label{sec:xray}

\begin{figure}
\centering
\includegraphics[scale=0.4]{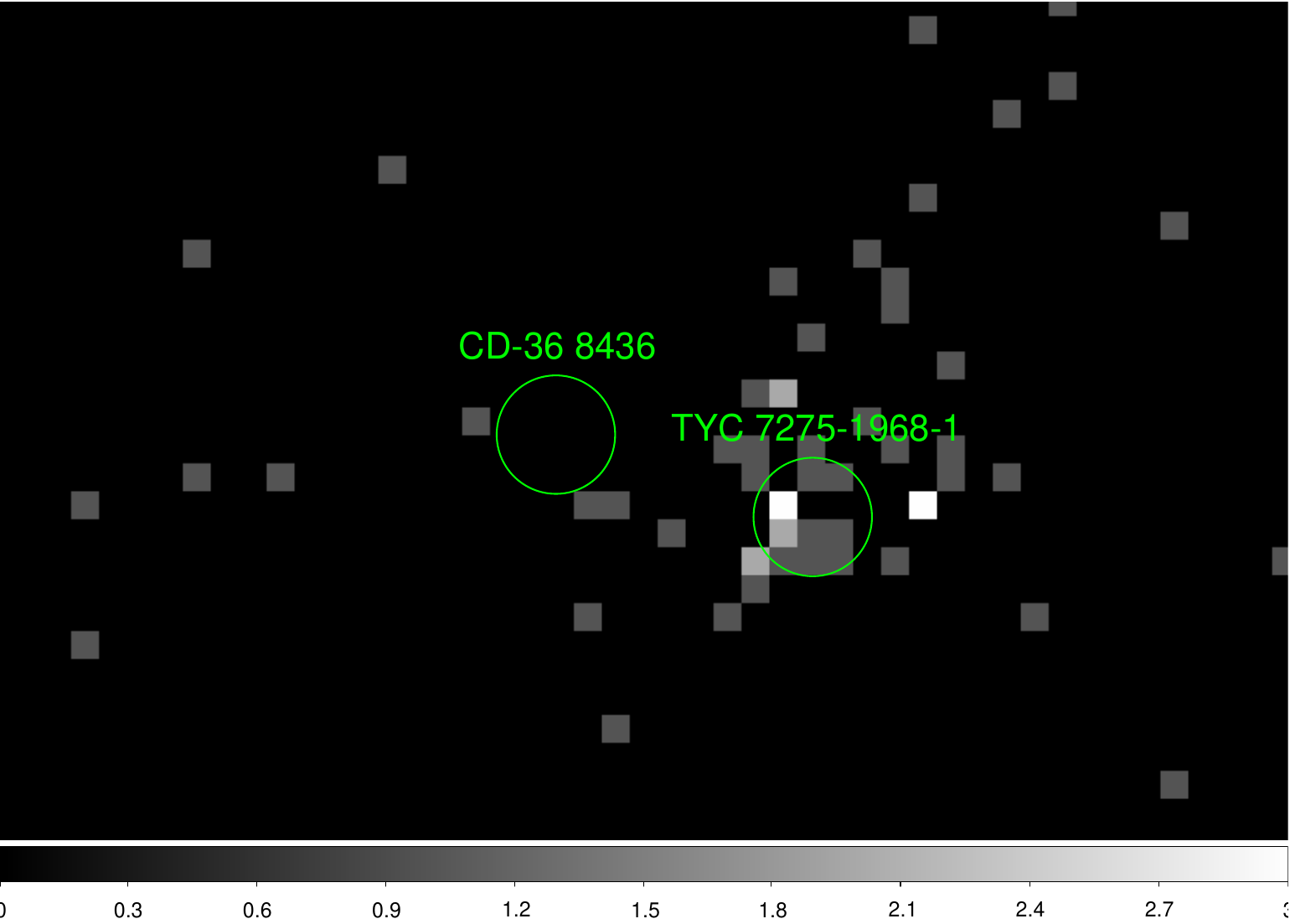} \\(a) \swift/XRT \\
\includegraphics[scale=0.4]{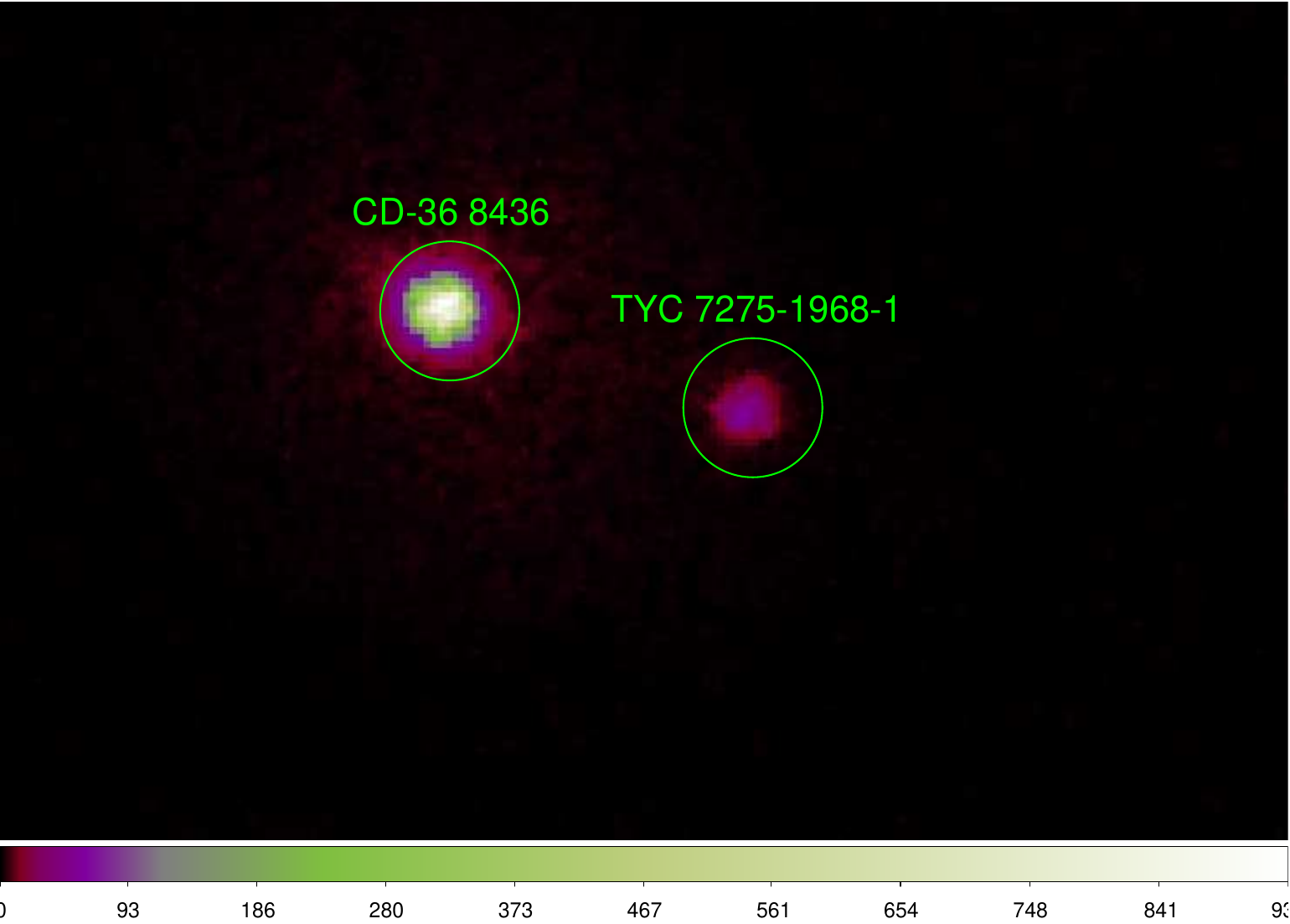} \\(b) \swift/UVOT\\
\caption{\swift\ XRT and UVOT images of the field of TYC~7275. The symbiotic star CD-36~8436 is not detected in X-rays, but it is brighter than TYC~7275 in the UV.}
\label{fig:FOV_CD-27}
\end{figure}

TYC~7275 is in the FoV of the \swift~X-ray Telescope (XRT) observations listed in Table~\ref{tab:log-observations}. The \swift\ catalog\footnote{see \url{https://www.swift.ac.uk/swift_live/index.php}} presents an X-ray source associated with CD~-36~8436 ($\alpha_{2000.0}$~=~13h 16m 01.37s, $\delta_{2000.0}$~=~-37$^{\circ}$ 00$^{'}$ 10.77$^{''}$), a symbiotic star located 22~arcsec from the coordinates of TYC~7275 ($\alpha_{2000.0}$~=~13h 15m 59.55s, $\delta_{2000.0}$~=~-37$^{\circ}$ 00$^{'}$ 17.73$^{''}$). Our analysis of the \swift\ data shows that the X-ray emission is associated with TYC~7275 and not with CD~-36~8436 (Figure~\ref{fig:FOV_CD-27}a). No X-ray emission from the symbiotic system is detected.

\begin{table}
\caption{Log of \swift\ observations. The cycle for each time interval of the observations are also given using period in Equation~\ref{tess_hjd_final}.} 
\smallskip
\centering
\label{tab:log-observations}
\begin{tabular}{ccccc} 
\hline\hline  
\noalign{\smallskip}
Date & ObsId &  XRT exp. time & UVOT exp. time & Cycles\\
 & & (ks) & (ks)  &  \\
\noalign{\smallskip}
\hline
\noalign{\smallskip}  
2013~Feb~09 & 00091459002 & 5.9 & 0.7 & 2.0\\  
2013~Mar~03 & 00091459004 & 0.1 & 0.1 & 0.1\\  
2013~Mar~09 & 00091459005 & 2.5 & 0.2 & 0.3\\ 
2013~Aug~30 & 00091748001 & 1.2 & 0.6 & 0.4\\ 
2013~Aug~31 & 00091748002 & 0.2 & 0.6 & 3.2\\ 
2014~Feb~25 & 00091748005 & 0.1 & 0.1 & 0.1\\ 
2014~Feb~28 & 00091748006 & 0.6 & 0.6 & 0.1\\ 
2014~Mar~11 & 00091748007 & 0.1 & 0.2 & 0.1\\ 
\hline \noalign{\smallskip} \end{tabular} 
\end{table}

We obtained the X-ray spectrum of TYC~7275 using the online \swift/XRT tool\footnote{See build \swift-XRT products in \url{https://www.swift.ac.uk/user_objects/}} provided by the UK \swift\ Science Data Center at the University of Leicester \citep{Evans_2009}. The spectrum is calculated combining all the available observations (see Table~\ref{tab:log-observations}). \swift\ detected the source with an average count rate of 0.0029~$\pm$~0.0005~counts~s$^{-1}$. 

Simultaneously with the X-ray Telescope, \swift\ observed the field of TYC~7275 with the UVOT telescope in the {\it event} mode with the UVM2 filter ($\lambda$2246~\AA, Figure~\ref{fig:FOV_CD-27}b). From the event files provided by the pipeline, we first filtered for bad events using the \texttt{uvotscreeen} script and then extracted the source events from a circular region of 5 arcsec radius, while background events were extracted from an off-source circular region with 20 arcsec radius. The light curves, with 60~s bins, were extracted using the \texttt{uvotevtlc} script. 

\section{Results}

\subsection{Timing analysis} \label{sec:time}

In order to search for the orbital period in the optical light curves, we used the 
Lomb-Scargle (LS) periodogram \citep{Lomb_1976,Scargle_1982} from the software {\it PERANSO} \citep{Paunzen_2016}. We also applied other methods such as the Generalized Lomb-Scargle (GLS), the Discrete Fourier Transform (DFT), the Date Compensated Discrete Fourier Transform (DCDFT), and the Fourier Analysis of Light Curves (FALC). The periods obtained using the different methods are consistent with each other. As the flux modulation has a sinusoidal shape, we opted to present the results from the LS method. Table~\ref{tab:periods-results} shows the period and its errors obtained using the data from individual and combined datasets. In general, we found the period of 0.1914035~d. The mean epoch of each survey in units of BJD$_{TDB}$ is also given in this table.

\begin{table}
\caption{Main period found in individual or combined datasets.}  
\smallskip
\centering
\label{tab:periods-results}
\begin{tabular}{ccccc} 
\hline\hline  
\noalign{\smallskip}
Surveys & Initial Time & Cycles$^{a}$ & Period & Period error \\
& (BJD$_{TDB}$) & & (d)  & (d)  \\
\noalign{\smallskip}
\hline
\noalign{\smallskip}  
ASAS-3 & 2452665.855859 & 5882 & 0.1914033 & 0.0000022\\ 
SuperWASP & 2453860.217572 & 1964 & 0.1914049 & 0.0000053\\ 
CRTS & 2453607.876401 & 4647 & 0.1914035 & 0.0000013\\ 
Gaia & 2456897.46333001 & 2420 & 0.1914035 & 0.0000053\\ 
ASAS-SN & 2455955.148353 & 11045 & 0.1914035 & 0.0000013 \\ 
TESS Sector 37 & 2459307.25853093 & 66 & 0.1914101 & 0.0000501\\
TESS Sector 64 & 2460042.12165516 & 60 & 0.1913940 & 0.0000633\\
All TESS &  2459307.25853093 & 1980 & 0.1914035 & 0.0000026\\
\hline
All data & $-$ & $-$ & 0.1914035 & 0.0000013\\ 
\hline \noalign{\smallskip} \end{tabular} \\
\noindent ${a}$ - the number of cycles enclosed in the time interval of the observations of each survey. 
\end{table}

\begin{figure}
\begin{center}
\includegraphics[scale=0.41,angle=0]{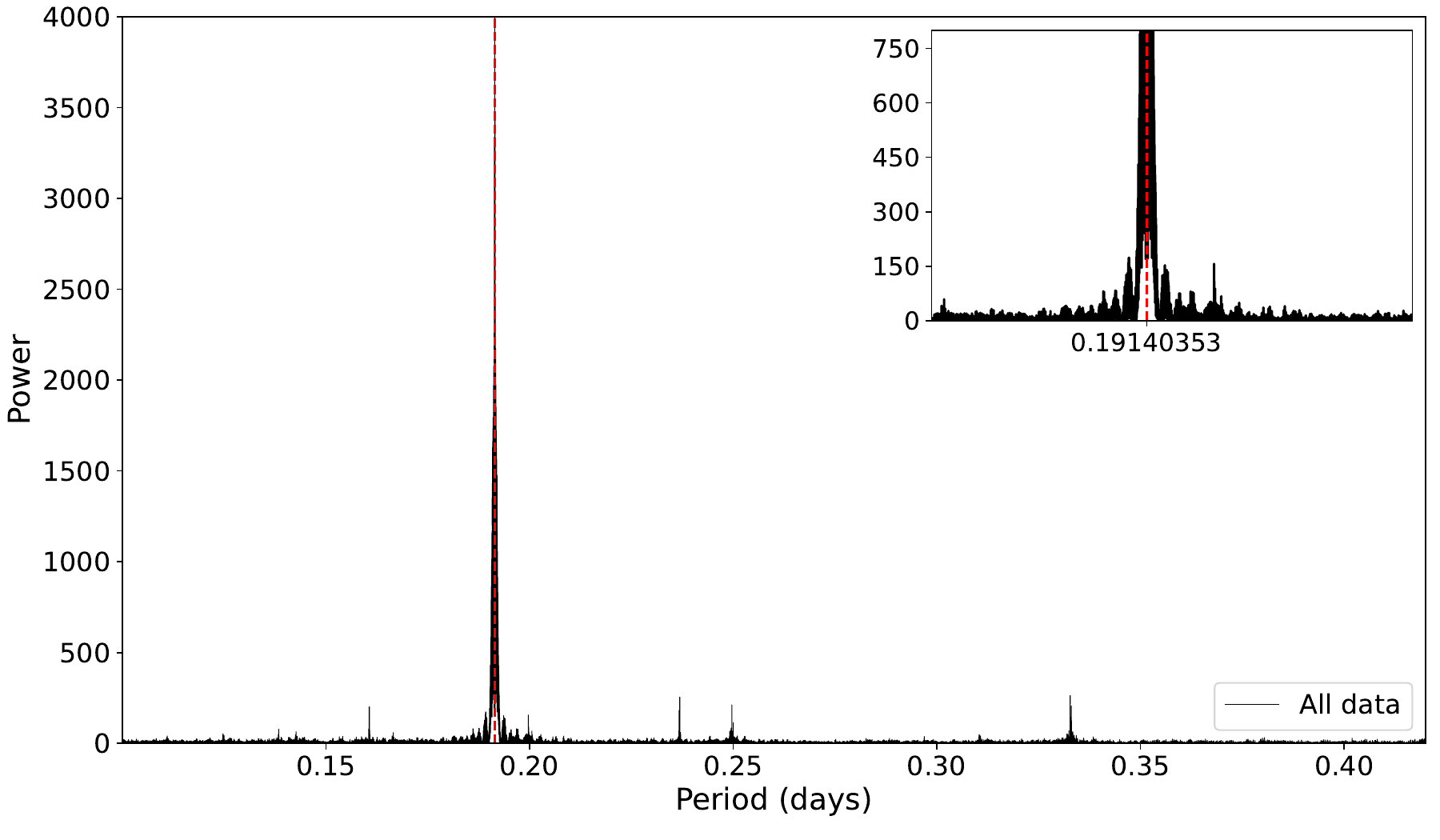}\\
\caption{The power spectrum of TYC~7275-1968-1 from \asass, \wasp, \crts, \gaia, \asas, and \tess\ combined dataset. The red line corresponds to period of 0.1914035~d.}
\label{fig:tess_lomb}
\end{center}
\end{figure}

\begin{figure}
\begin{center}
\includegraphics[scale=0.45,angle=0]{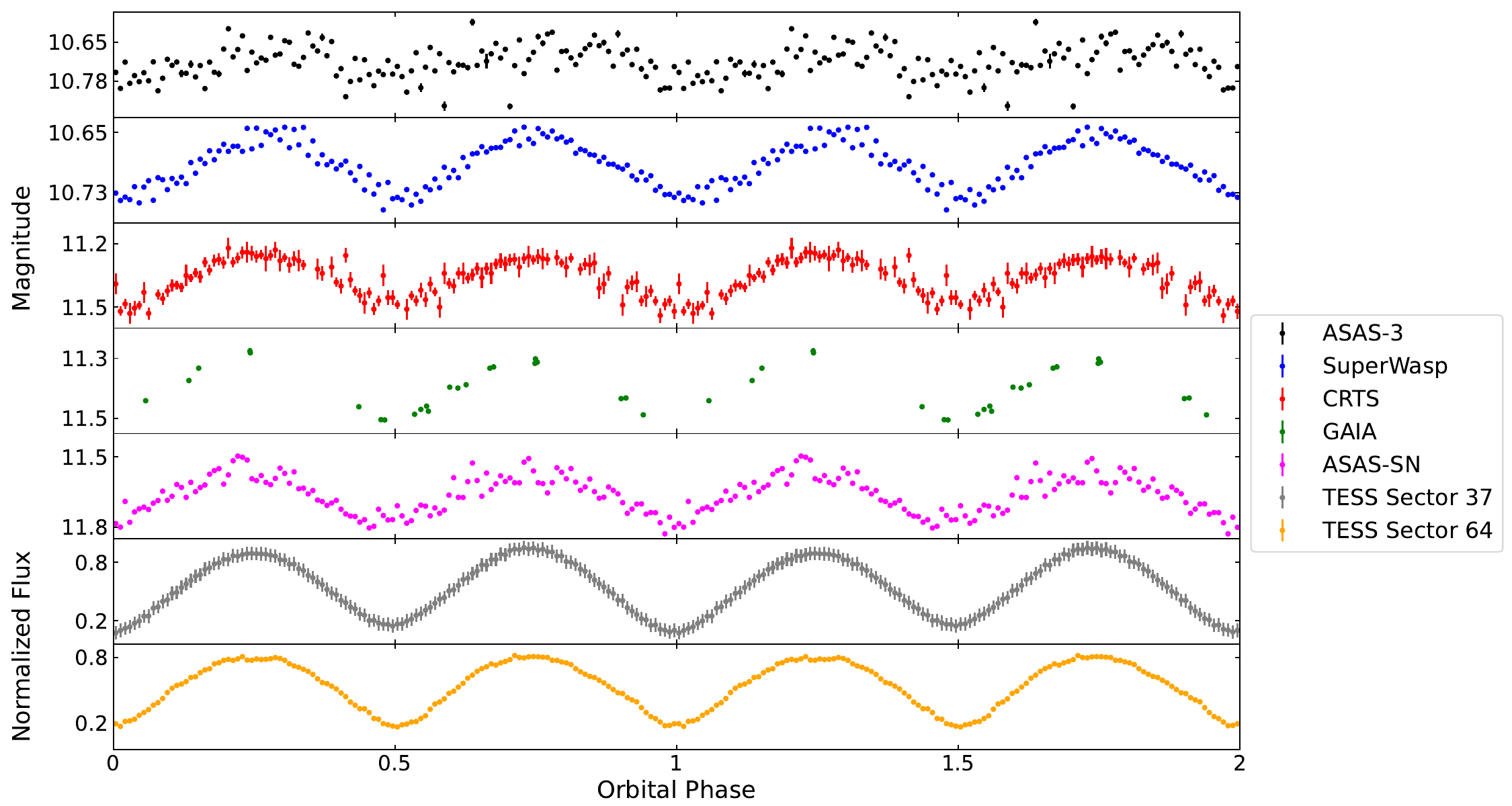} \\
\caption{Folded light curves at the 0.3828071~d period and T$_{0}$ of the ephemeris for \asass, \wasp, \crts, \gaia, \asas, and \tess-Sector~37 and -Sector~64 data.}
\label{fig:tess_light}
\end{center}
\end{figure}

Using all data combined, we found a strong signal on the power spectrum at 0.1914035~$\pm$~0.0000013~d  (see Figure~\ref{fig:tess_lomb}). 
This period is the first harmonic of the orbital period of 0.3828071~$\pm$~0.0000026~d. The light curves of all data sets folded in this period and binned at 120 bins/cycle are shown in Figure~\ref{fig:tess_light}, where we used the ephemeris:

\begin{equation}    T_{0}~=~BJD_{TDB}~2459315.09206104~+~0.3828071E(0.0000026),
\label{tess_hjd_final}
\end{equation}
\noindent where the phase zero was visually defined inspecting the phase-folded of \tess-Sector 37.

\begin{figure}
\begin{center}
\includegraphics[scale=0.4,angle=0]{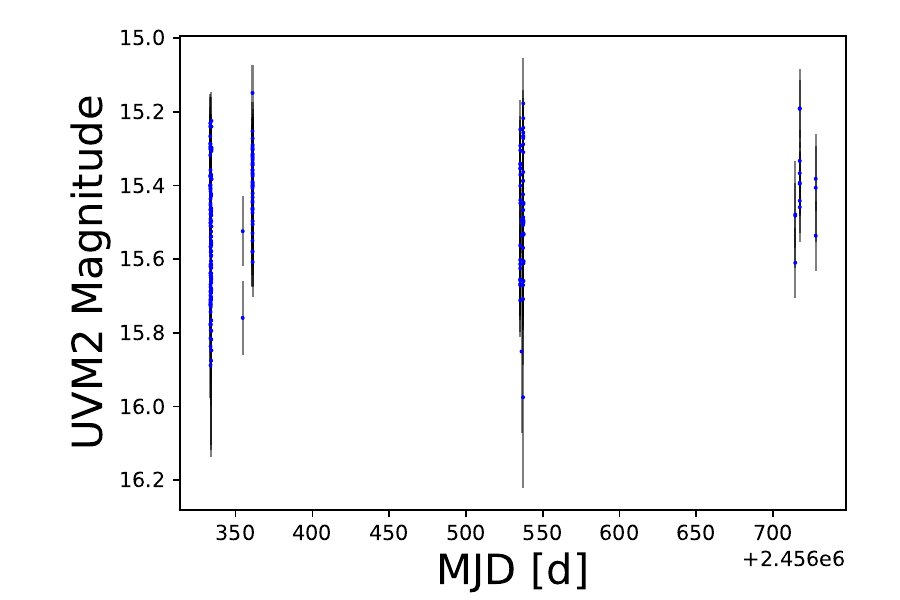}
\includegraphics[scale=0.4,angle=0]{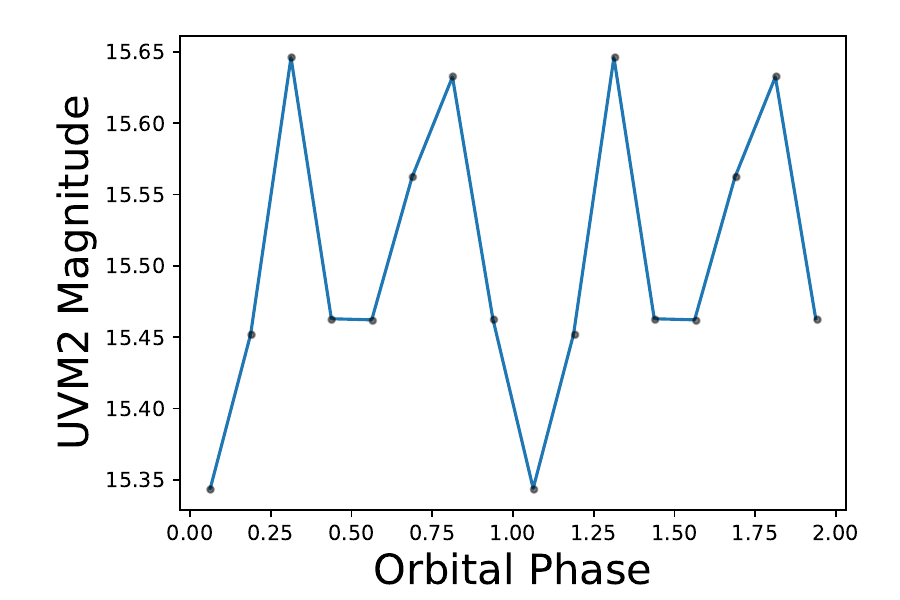} 
\caption{\swift/UVOT data of TYC~7275-1968-1. The light curve on left panel and folded on 0.3828071 days orbital period and T$_{0}$ of the ephemeris on right panel.}
\label{fig:uvm2}
\end{center}
\end{figure}

In the \swift/UVOT UVM2 filter, TYC~7275 is much fainter than the symbiotic star CD~-36~8436 (see Figure~\ref{fig:FOV_CD-27}b), and because of the scarcity of UV signal we did not search for periodicities in the UV data. Figure~\ref{fig:uvm2} shows the UVM2 light curve and folded light curve in the orbital period, where both primary and secondary eclipses are detected.

\subsection{XRT spectral analysis}
\label{sec:xrays}

\begin{figure}
\begin{center}
\includegraphics[scale=0.5,angle=0]{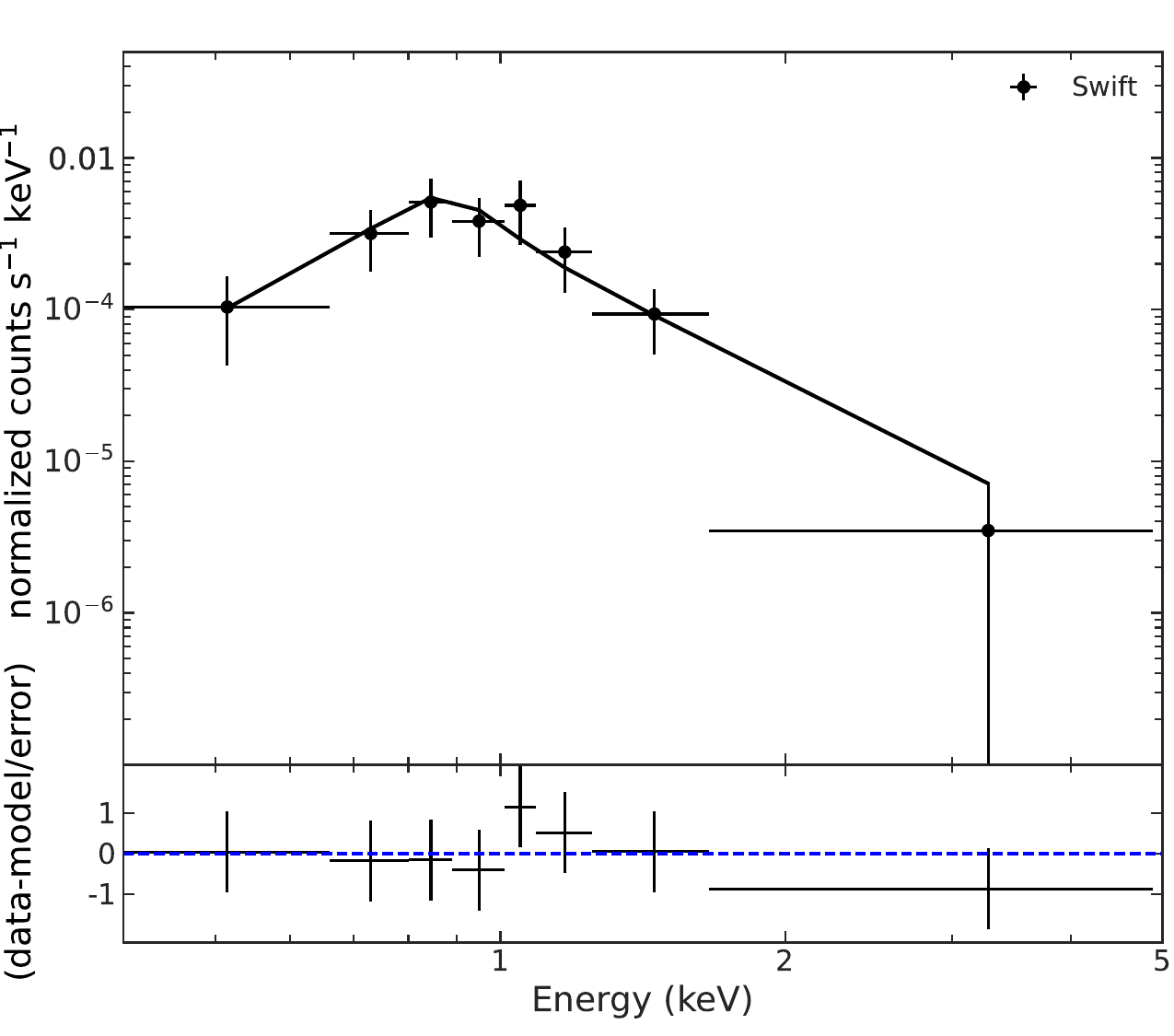}
\caption{\swift/XRT X-ray spectrum of TYC~7275-1968-1. The solid line indicates the best fit model.}
\label{fig:xray}
\end{center}
\end{figure}

In spite of this low counts, we performed a simple spectral fit using a model consisting of an optically thin thermal plasma (\textit{APEC}) modified by an absorber (\textit{tbabs}). The spectrum is presented in Figure~\ref{fig:xray}. We obtained the absorbing column density of N(H)~=~0.17$^{+0.2}_{-0.1}$~$\times$~10$^{22}$ cm$^{-2}$ and kT~=~0.8$^{+0.9}_{-0.1}$~keV. The metal abundance was 0.23$^{+0.67}_{-0.01}$ \citep{Wilms_2000}. Taking into account the above model, we obtained an unabsorbed X-ray ﬂux of 1.2$^{+0.1}_{-0.2}$~$\times$10$^{-13}$~erg~s~$^{-1}$~cm$^{-2}$, which converts to a luminosity of 1.4$^{+0.1}_{-0.2}$~$\times$~10$^{31}$~erg~s~$^{-1}$ at a distance of 317$^{+2}_{-2}$~pc as given by Gaia (EDR3) \citep{Bailer-Jones_2021}.

\section{Discussion and conclusions} \label{sec:discussion}

In this work, we present an analysis of X-ray and optical data from \swift, \asass, \wasp, \crts, \gaia, \asas, and \tess\ of a recently classified red nova progenitor, the contact binary system TYC~7275. The current study shows that the X-ray emission detected by \swift,
as well as the optical variability measured by \tess, belongs to this object and not to the symbiotic star CD~-36~8436 as previously registered, which is located 22~arcsec from TYC~7275.

\begin{figure}
\begin{center}
\includegraphics[scale=0.5,angle=0]{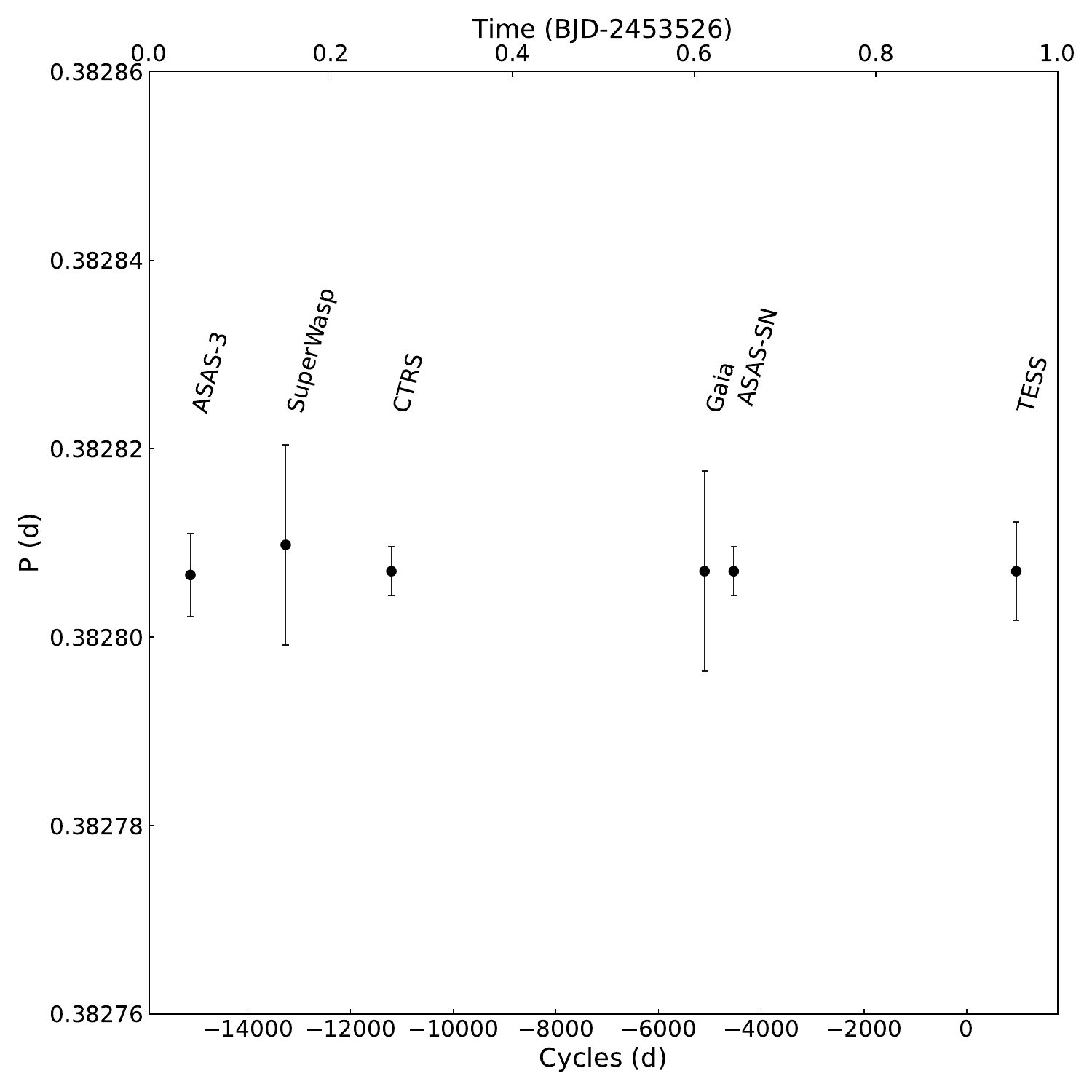}
\caption{Orbital period versus the mean epoch of each dataset. The period is constant along of the 22 years of optical observations.} 
\label{fig:period_time}
\end{center}
\end{figure}

The optical photometry data were used to improve the orbital period of estimate to 0.3828071(26)~d.
This orbital period and the low mass ratio indicate that the system can be orbitally unstable and prone to a merger event \citep{Wadhwa_2022a,Wadhwa_2022b}. However, we did not detected a significant period variation during the 22 years of optical observations, as shown in Figure~\ref{fig:period_time} and Table~\ref{tab:periods-results}. 

The temperature of the hot component can be estimated using \gaia\ intrinsic color and interstellar extinction \citep{bakics_2022}.
Adopting a \gaia\ color $Bp-Rp$ of 0.865942 mag and an extinction, A(V), of 0.1~mag, we obtained a temperature of 5600~K, which is similar to the models with unspotted and spotted solutions proposed by \cite{Wadhwa_2022a}.

The X-ray spectrum from \swift\ can be adequately modeled with an absorbed optically thin thermal plasma with a temperature of $kT$~=~0.8$^{+0.9}_{-0.1}$. This X-ray emission probably originates in a corona around the contact binary. The X-ray luminosity measured by \swift\ is 1.4$^{+0.1}_{-0.2}$~$\times$~10$^{31}$~erg~s$^{-1}$ and is higher than that reported by \cite{Schmitt_2001} and \cite{Wadhwa_2022a} of 1.7$\times$10$^{30}$~ergs~s$^{-1}$ from \rosat\ observations. This change in luminosity suggests that the components of the binary might be magnetically active \citep{Applegate_1992}.

\section{Acknowledgments}

IJL, GJML, and NEN acknowledge support from grant ANPCYT-PICT 0901/2017. GJML and NEN are members of the CIC-CONICET (Argentina). ACM thanks the Brazilian \textit{Conselho Nacional de Desenvolvimento Científico e Tecnológico} -- CNPq (Proc: 382618/2021-1). ASO acknowledges São Paulo Research Foundation (FAPESP) for financial support under grant \#2017/20309-7. CVR thanks CNPq (Proc: 310930/2021-9). NP thanks the Coordenação de Aperfeiçoamento de Pessoal de Nível Superior -- Brazil (CAPES) for the financial support under grant 88887.823264/2023-00. This research made use of Lightkurve, a Python package for Kepler and TESS data analysis (Lightkurve Collaboration, 2018). We acknowledge \asass, \wasp, \crts, \gaia, \asas\ and Astropy Collaboration. The authors are also grateful to the TESS High Level Science Products (HLSP) produced by the Quick-Look Pipeline (QLP) at the TESS Science Office at MIT, which are publicly available from the Mikulski Archive for Space Telescopes (MAST). Funding for the TESS mission is provided by NASA's Science Mission directorate.







\bibliography{manuscript}{}
\bibliographystyle{aasjournal}


\end{document}